\RequirePackage{fixltx2e}
\documentclass[
    10pt,
    ,aps
    ,pop%,rmp%,pra%,pop
    ,oneside
    ,nobibnotes
    ,showpacs
    ,twocolumn
]{revtex4-1}

\usepackage[
    colorlinks
    ,unicode
    ,pdfpagemode=UseOutlines% (show bookmarks)
    ,pdfdisplaydoctitle=true% display document title instead of file
    ,pdfpagetransition=Wipe%Dissolve
]{hyperref}
    \hypersetup{
        pdftitle={All About Helicons}
        ,pdfauthor={Igor A. Kotelnikov}
        ,pdfsubject={Draft, \today}
        ,pdfkeywords={helicons, whistlers, low hybrid resonance, gas-dynamic trap}
    }
\usepackage{graphicx}\graphicspath{{}{Graph/}{Graph/seminar/}}
\usepackage{bookmark}

\usepackage{amsmath,amssymb,bm}
\allowdisplaybreaks[1]

\usepackage{ifluatex,ifxetex}
\ifnum 0\ifxetex 1\fi\ifluatex 1\fi>0
    \usepackage{fontspec}
    \defaultfontfeatures{Ligatures={TeX},Renderer=Basic}

    \AtBeginDocument{\def\perp{\mathrel\bot}}

    \usepackage[math-style=ISO,bold-style=ISO]{unicode-math}
    \mathversion{cambria}
    \renewcommand{\bm}{}
\else
    \usepackage[T1]{fontenc} % UTF-8: PDFLaTeX
    \usepackage[utf8]{inputenc} % UTF-8: PDFLaTeX
\fi

\usepackage[english]{babel}
    \renewcommand{\vec}[1]{\mathbf{#1}}

\usepackage{tensor}

\makeatletter
\@ifpackageloaded{unicode-math}{%
  \DeclareRobustCommand{\tensor}[1]{\overleftrightarrow{{\mathbf{#1}}}}%
  \PackageInfo{handy}{\string\tensor\space has been declared using ^^J\space\space \string\overleftrightarrow\space from unicode-math package}%
}{%
  \@ifpackageloaded{accents}{%	
    \DeclareRobustCommand{\tensor}[1]{\accentset{\leftrightarrow}{{\mathbf{#1}}}}%
    \PackageInfo{handy}{\string\tensor\space has been declared using ^^J\space\space \string\accentset\space from in accents package}%
  }{%
    \@ifpackageloaded{revsymb4-1}{%
      \DeclareRobustCommand\tensor[1]{\@ontopof{\mathbf{#1}}{\leftrightarrow}{1.15}\mathord{\box2}}
      \PackageWarningNoLine{handy2}{\string\tensor\space has been taken from revsymb4-1 package.^^J You are recommended to load `accents' package before me}%
    }{
    \@ifpackageloaded{revsymb}{%
        \PackageWarningNoLine{handy2}{\string\tensor\space has been taken from revsymb package.^^J You are recommended to load `accents' package before me}%
      }{%
        \ifx\undefined\overset
        \else
          \providecommand{\tensor}[1]{\overset{\leftrightarrow}{\vec{#1}}}
          \PackageWarningNoLine{handy2}{\string\tensor\space has been declared using ^^J\space\space \string\overset\space from amsmath package.^^J You are recommended to load `accents' package before me}%
        \fi
      }%
    }%
  }%
}%
\makeatother
\begin{document}

\def\perp{\mathrel\bot}

%%\title{The coalescence phenomena and the Golant-Stix criterion}
%%\title{Coalescence phenomena and a limiting density in the helicon plasma sources}
%%\title{Coalescence phenomenon and the density limit in the helicon plasma sources}
\title{On the density limit in the helicon plasma sources}
\author{Igor A. Kotelnikov}
    \email{I.A.Kotelnikov@inp.nsk.su}
    \affiliation{Budker Institute of Nuclear Physics, Novosibirsk, Russia}
    \affiliation{Novosibirsk State University, Novosibirsk, Russia}

%%%% PACS numbers
\pacs{
    % ==========================
    %http://publish.aps.org/PACS
    % ==========================
    %52.35.Fp; %	Electrostatic waves and oscillations (e.g., ion-acoustic waves)
    52.35.Hr; %	Electromagnetic waves (e.g., electron-cyclotron, Whistler, Bernstein, upper hybrid, lower hybrid)
    52.50.Qt; %	Plasma heating by radio-frequency fields; ICR, ICP, helicons
    52.50.Sw; %	Plasma heating by microwaves; ECR, LH, collisional heating
}

\begin{abstract}
    %The accessibility of the lower hybrid resonance zone in a radially inhomogeneous plasma column is critically revisited. A new derivation of the Golant-Stix criterion is given. It yields an exact but rather simple expression for a critical value of the parallel refractive index $N_{\|}=k_{\|}c/\omega$ with  $N_{\|}^{2} > \omega_{ce}\omega_{ci}/(\omega_{ce}\omega_{ci}-\omega^{2})$ assuming that the lower hybrid resonant zone is accessible for a radiation launched from outside of the plasma column. A feasibility of the lower hybrid resonant heating of a dense plasma core is discussed and an optimal magnetic field is found. The Shamrai-Taranov criterion is also discussed and its connection with the Golant-Stix criterion is elucidated. It is shown that both criteria are related to the effect of coalescence of the helicon and Trivelpiece-Gould waves but are intended for different frequency ranges.

    Existence of the density limit in the helicon plasma sources is critically revisited. The low- and high-frequency regimes of a helicon plasma source operation are distinguished. In the low-frequency regime with $\omega < \sqrt{\omega_{ci}\omega_{ce}}$ the density limit is deduced from the Golant-Stix criterion of the accessibility of the lower hybrid resonance. In the high-frequency case, $\omega > \sqrt{\omega_{ci}\omega_{ce}}$, an appropriate limit is given by the Shamrai-Taranov criterion. Both these criteria are closely related to the phenomenon of the coalescence of the helicon wave with the Trivelpiece-Gould mode. We argue that theoretical density limits are not achieved in existing devices but might be met in the future with the increase of applied rf power.
\end{abstract}

\maketitle

\section{Introduction}
\label{1}

One of the most challenging problems in the theory of helicon plasma sources is a supposed existence of the plasma density limit \cite{Chen155, Ellingboe1996PoP_3_2797, ShamraiTaranov1996PSSciTech_5_474, Lafleur+2011JPhysD_44_055202, Shinohara+2013FST_63_164}. For the helicon plasma sources, it is conventional to consider the density of the order of $10^{12}\;\text{cm}^{-3}$ as very high, but preproduction of plasma for fusion devices needs the density of the order of $10^{14}\;\text{cm}^{-3}$, at least. For this reason, in this paper we critically revisit the foundations of the theory of helicon heating, assuming that rf power is sufficiently large for a plasma source to operate in the so called W-mode (helicon-Wave mode) as explained in Refs.~\cite{EllingboeBoswell1996PoP_3_2797, Chabert+2011physics}.

We distinguish a low-frequency and a high-frequency regimes of operation of a helicon source where the frequency $\omega$ is respectively smaller and larger than the hybrid cyclotron frequency $\sqrt{\omega_{ce}\omega_{ci}}$.

The low-frequency regime ($\omega_{ci} < \omega <\sqrt{\omega_{ce}\omega_{ci}}$) is characterized by existence of the lower hybrid resonance. We reexamine the accessibility condition of the resonance in a radially inhomogeneous cylindrical plasma column, which is uniform along its axis. This condition is known as the Golant-Stix criterion \cite{Golant1972SovPhysTechPhys_16_1980, Stix1962, Stix1992}. We provide a new derivation of this criterion, which reveals its connection to the effect of the wave coalescence. We  find that a density limit indeed exist in this regime. However it has not a feature of a threshold since the limiting density depends on the value of the longitudinal refractive index $N_{\|}=k_{\|}c/\omega$ (the larger $N_{\|}$, the larger the limit), whereas the spectrum of the plasma oscillations excited by an antenna is usually quite wide. Therefore, speaking about a limiting density, we imply a value of $N_{\|}$, which corresponds approximately to the maximum in the absorption spectrum of the antenna.

In the high-frequency regime ($\sqrt{\omega_{ce}\omega_{ci}} < \omega < \omega _{ce}$), hybrid resonances are not available, and the density limit occurs because of the coalescence of the helicon and Trivelpiece-Gould (TG) waves. Corresponding density limit is given by the Shamrai-Taranov criterion \cite{ShamraiTaranov1996PSSciTech_5_474}.

The paper is organized as follows. The main equations are reviewed in Secs.~\ref{s2} and~\ref{s3}. The Golant-Stix criterion is discussed in Sec.~\ref{s4}. A limiting plasma density in the low-frequency mode of operation of the helicon source is found in Sec.~\ref{s5}. An optimal magnetic field is evaluated in Sec.~\ref{s6}. The high-frequency mode of the helicon source is considered in  Sec.~\ref{s7}, where a new simple derivation of the Shamrai-Taranov criterion is given. Finally, in Sec.~\ref{s8} we discuss how the low-frequency regime matches the high-frequency regime of operation.

\section{Dispersion Equation}
\label{s2}

We consider a \emph{simple plasma} consisting of the electrons and a single kind of ions.  In the approximation of cold collisionless plasma, both plasma species are characterized by the two quantities each, the Langmuir frequency $\omega_{ps}=\sqrt{4\pi e_{s}^{2}n_{s}/m_{s}}$ and the Larmor frequency $\Omega_{s}=e_{s}B/m_{s}c$ with the subscript $s=e$ standing for the electrons and $s=i$ for the ions. Assuming that the magnetic field $\vec{B}$ is directed along the axis $z$ of the axial symmetry of the plasma column, the permittivity tensor   reads
    \begin{equation}
    \label{2.3:1}
    \tensor{\vec{\varepsilon}}
    =
    \begin{pmatrix}
      \varepsilon & ig & 0\\
      -ig & \varepsilon & 0\\
      0 & 0 & \eta
    \end{pmatrix}
%    =
%    \begin{pmatrix}
%      S & -iD & 0\\
%      iD & \varepsilon & 0\\
%      0 & 0 & P
%    \end{pmatrix}
%    .
    ,
    \end{equation}
where
    \begin{subequations}
    \label{2.3:2}
    \begin{gather}
    \label{2.3:2a}
    \varepsilon
    =
    \frac{\varepsilon_{+}+\varepsilon_{-}}{2}
    = 1
      - \frac{
        \omega_{p}^{2}
        \left(\omega^{2}+\Omega_{e}\Omega_{i}\right)
      }{
        \left(\omega^{2}-\Omega_{e}^{2}\right)
        \left(\omega^{2}-\Omega_{i}^{2}\right)
      }
    ,
    \\
    %\label{2.3:2b}
    g
    =
    \frac{\varepsilon_{+}-\varepsilon_{-}}{2}
    =
    -
    \frac{
        \omega_{p}^{2} \omega
        \left(\Omega_{e}+\Omega_{i}\right)
    }{
        \left(\omega^{2}-\Omega_{e}^{2}\right)
        \left(\omega^{2}-\Omega_{i}^{2}\right)
    }
    ,
    \\
    %\label{2.3:2c}
    \eta
    =
    1
    -
    \frac{\omega_{p}^{2}}{\omega^{2}}
    ,
    \\
    \label{2.3:2d}
    \varepsilon_{\pm}
    =
    1
    -
    \frac{
        \omega_{p}^{2}
    }{
        \left(\omega\mp\Omega_{e}\right)\left(\omega\mp\Omega_{i}\right)
    }
    ,
    \\
    \label{2.3:2f}
    \omega_{p}^{2} \equiv \omega_{pe}^{2}+\omega_{pi}^{2}
%    ,
%    \\
%    \label{2.3:2g}
%    \omega_{ps}^{2}=\frac{4\pi e_{s}^{2} n_{s}}{m_{s}},
%    \\
%    \label{2.3:2h}
%    \Omega_{s}=\frac{e_{s}B}{m_{s}c}
    .
    \end{gather}
    \end{subequations}
These expressions are derived from the well-known formulas
    \begin{gather*}
    \label{2.3:6}
      \varepsilon = 1 - \sum_{s} \frac{\omega_{ps}^{2}}{\omega^{2}-\Omega_{s}^{2}}
      ,
      \\
      g =  \sum_{s} \frac{\Omega_{s}}{\omega}\frac{\omega_{ps}^{2}}{\omega^{2}-\Omega_{s}^{2}}
      ,
      \\
      \varepsilon_{\pm} = \varepsilon \mp g
      ,
    \end{gather*}
using the quasi-neutrality condition
    \begin{gather}
    \label{2.3.7}
    \omega_{pe}^{2}\Omega_{i}+\omega_{pi}^{2}\Omega_{e}=0
    .
    \end{gather}
Since $\Omega_{e}<0$, below we also use alternative notations
    \[
    %\label{2.3:8}
    \omega_{ce} = -\Omega_{e} = \frac{|e|B}{m_{e}c}
    ,
    \qquad
    \omega_{ci} = \Omega_{i} = \frac{e_{i}B}{m_{i}c}
    \]
for the cyclotron frequencies when it is more convenient to operate with designedly positive frequencies.

\begin{figure}[!t]
  \parbox{0.49\columnwidth}{
    % See PlasmaBook2, ch.20
    \includegraphics[width=0.49\columnwidth]{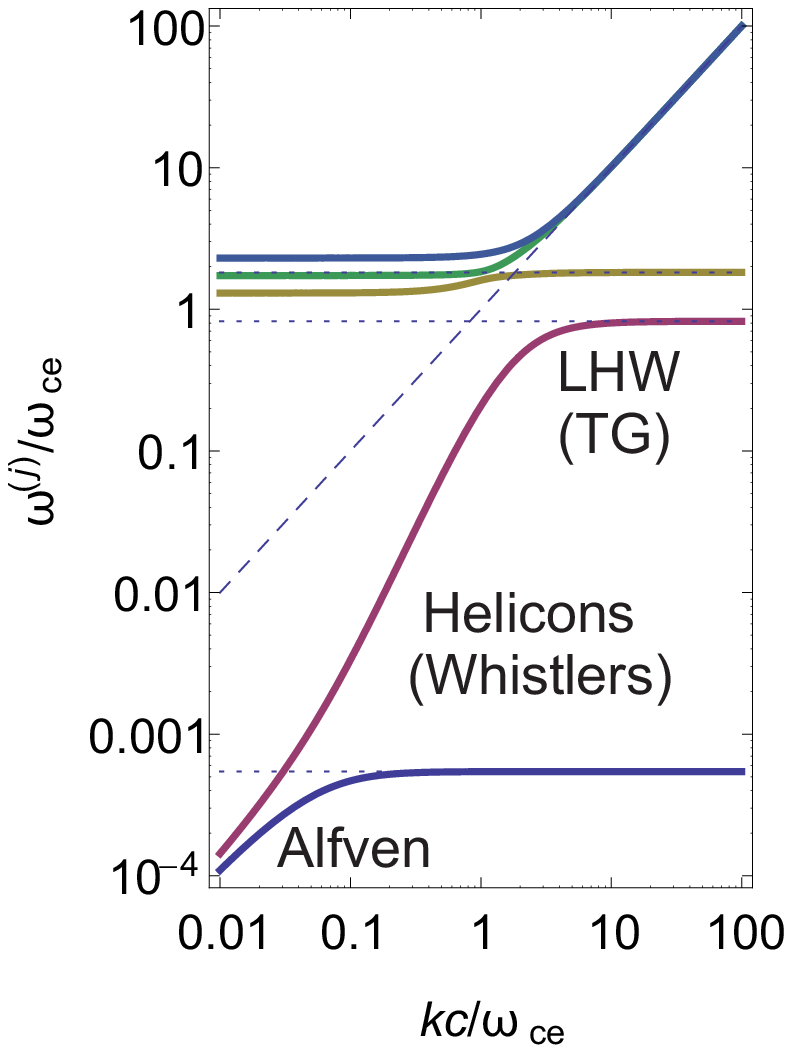}
  }\hfill
  \parbox{0.49\columnwidth}{
    \includegraphics[width=0.49\columnwidth]{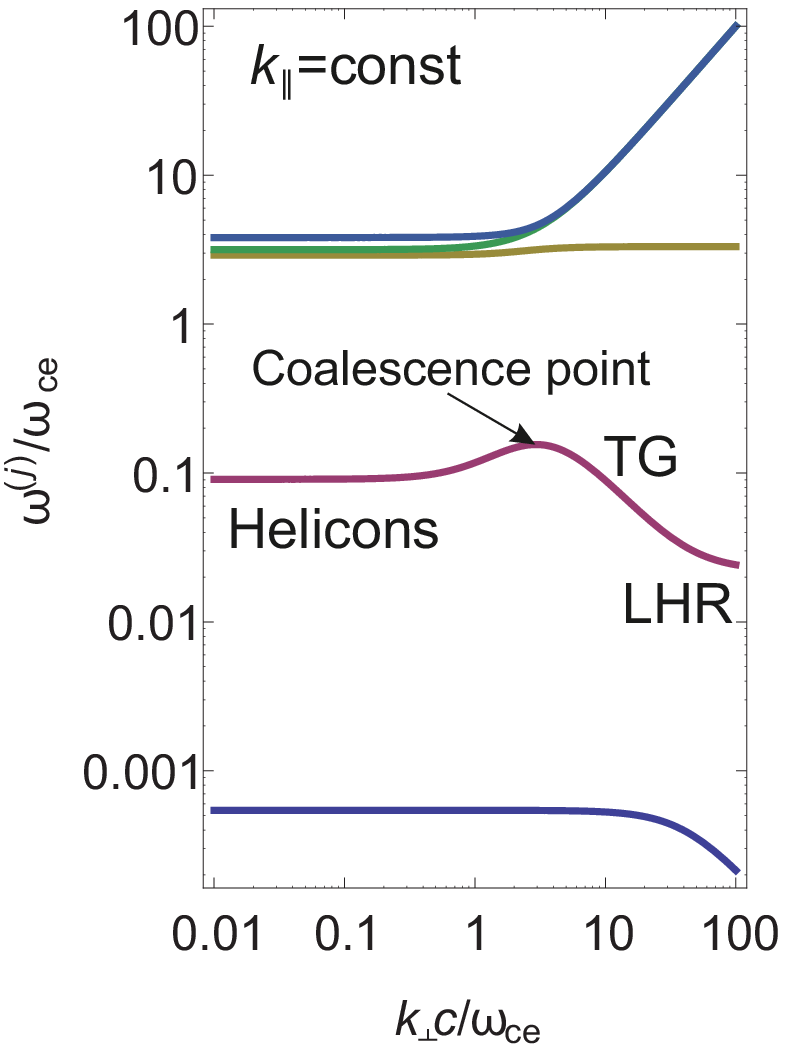}
  }\\
  \parbox[t]{0.49\columnwidth}{
    \caption{
        (Color online)
        Dispersion of the cold plasma waves at fixed angle of propagation: $\theta=\pi/6$, $\omega_{p}^{2}/\omega_{ce}^{2}=3$. The terms helicons (whistlers), Trivelpiece-Gould (TG) and low hybrid resonance (LHR) designate different parts of the same branch $\omega^{(2)}$ of the dispersion relation (purple curve).
    }\label{fig:Spectrum}
  }\hfill
  \parbox[t]{0.49\columnwidth}{
    \caption{
        (Color online)
        Dispersion of the cold plasma waves at fixed $k_{\|}=\omega _{ce}/c$, $\omega_{p}^{2}/\omega_{ce}^{2}=10$. The helicons and TG modes are separated by a coalescence point. Other branches: Alfven wave (blue lower curve); slow extraordinary, upper hybrid wave (yellow);  ordinary wave (green); fast extraordinary wave (upper blue).
    }\label{fig:Spectrum-2}
  }
\end{figure}

A dispersion equation is obtained from the wave equation
    \begin{gather*}
    %\label{2.3:11}
%    N^{2}\vec{E}-\vec{N}\left(\vec{N}\cdot\vec{E}\right)
%    -\tensor{\bm\epsilon}\cdot\vec{E}
    \tensor{\bm\epsilon}{}\cdot\vec{E}
    +\vec{N}\left(\vec{N}\cdot\vec{E}\right)
    -N^{2}\vec{E}
    =0
    ,
    \end{gather*}
written in the Fourier domain with $\vec{N}=c\vec{k}/\omega$ denoting the vector of the refractive index. The same equation in the matrix form reads
    \begin{gather*}
    %\label{2.3:12}
    \begin{pmatrix}
      \varepsilon - N_{\|}^{2} & -ig & N_{\bot}N_{\|} \\
      ig & \varepsilon - N^{2} &  0 \\
      N_{\|}N_{\bot} & 0 & \eta - N_{\bot}^{2}
    \end{pmatrix}
    \begin{pmatrix}
      E_{x} \\ E_{y} \\ E_{z}
    \end{pmatrix}
    =0
    ,
    \end{gather*}
where $N_{\bot}=N_{x}=N\sin\theta $, $N_{\|}=N_{z}=N\cos\theta $, and $N_{y}$ is assumed to be zero. Equating the determinant of this equation to zero yields the dispersion equation
    \begin{gather}
    \label{2.3:13}
    \mathbb{A} N_{\bot}^4-\mathbb{B}N_{\bot}^{2}+\mathbb{C}=0
    ,
    \end{gather}
where
    \begin{gather*}
    %\label{2.3:15}
    \mathbb{A}
    =
    \varepsilon
    ,\\
    \mathbb{B}
    =
    \varepsilon_{+} \varepsilon_{-}
    + \eta \varepsilon
    - \varepsilon N_{\|}^{2} - \eta N_{\|}^{2}
    ,\\
    \mathbb{C}
    =
%    \eta N_{\|}^4
%    - 2 \eta \varepsilon N_{\|}^{2}
%    + \eta \varepsilon_{+} \varepsilon_{-}
%    =
    \eta
    \left(N_{\|}^{2}-\varepsilon_{+}\right)
    \left(N_{\|}^{2}-\varepsilon_{-}\right)
    .
    \end{gather*}
It is quadratic regarding $N_{\bot}^{2}$ and consequently has two solutions
    \begin{gather}
    \label{2.3:14}
    N_{\bot\pm}^{2}=(\mathbb{B} \pm\sqrt{\mathbb{B}^{2}-4\mathbb{A}\mathbb{C}})/2\mathbb{A}
    .
    \end{gather}
However at a given $k$ and $\theta$ the same dispersion equation yields 5 eigenfrequencies $\omega^{(j)}$ as shown in Fig.~\ref{fig:Spectrum}. The two solutions \eqref{2.3:14} mean that only 2 eigenmodes at most can simultaneously propagate at a given frequency $\omega$. They differ by the magnitude of $k$ and their polarizations. We will focus on the eigenfrequency $\omega^{(2)}$ which is the second by the magnitude of $\omega$; it is shown in purple color in Fig.~\ref{fig:Spectrum}.

In the region of interest $\omega \sim \sqrt{\omega_{ce}\omega _{ci}}$ near the lower hybrid frequency $\omega_{\text{LH}}$ (see below) the larger root $N_{\bot+}$ corresponds to a lower hybrid wave, which in last years is also called the Trivelpiece-Gould (TG) wave  \cite{TrivelpieceGould1959JAP_30_1784}. The smaller root $N_{\bot-}$ corresponds to the helicons which were called whistlers in the past. However, all these waves lie on the same curve on the graph of the frequency $\omega^{(2)}(k,\theta)$ versus the wave number $k$ if a value of the angle $\theta$ between the direction of the wave vector $\vec{k}$ and magnetic field $\vec{B} $ is fixed as shown in Fig.~\ref{fig:Spectrum}. The lowest frequency portion of the purple curve, where $\omega^{(2)}\propto k$, corresponds to the fast magnetosonic wave. When $\omega^{(2)}>\omega_{ci}$, it transforms into the helicons with the dispersion $\omega^{(2)} \propto k^{2}\cos\theta$. Finally, as we approach the lower hybrid frequency $\omega_\text{LH}$, which in a dense plasma is approximately equal to $\omega_\text{LH}\approx \sqrt{\omega_{ce}\omega_{ci}}$ the helicons are transformed into the Trivelpiece-Gould waves; in this region $\omega\approx \omega_\text{LH}$ and the frequency is almost independent of $k$.

\section{Singular points}
\label{s3}

Standard approach to the study of wave propagation in a cold magnetized plasma includes a search of singular points. Most textbooks distinguish two kinds of such points, namely the plasma (hybrid) resonances, where $N\to\infty$, and the cutoffs, where $N\to0$; see e.g. \cite{Shafranov1963VTP3(eng), Akhiezer+1974(eng), Stix1992}. This exhausts the list of singular points in a cold plasma at a fixed angle $\theta$ between the wave vector $\vec{k}$ and the magnetic field $\vec{B}$. When considering the propagation of waves in a cylinder with a radially inhomogeneous profile of the plasma density, we should rather assume that a fixed quantity is the longitudinal component of the wave vector $k_{\|}=k\cos\theta$, or, equivalently, the longitudinal component of the refractive index $N_{\|} = k_{\|}c/\omega$. For a fixed $N_{\|}$, one more type of the singular points appears, namely the point of coalescence, where the two roots \eqref{2.3:14} of the dispersion equation merge. The coalescence point of the helicon and TG waves is readily seen in Fig.~\ref{fig:Spectrum-2}. In contrast to Fig.~\ref{fig:Spectrum}, the curve $\omega^{(2)}(k_{\|},k_{\bot})$ is not a monotonically rising function of $k_{\bot}$, and, at a fixed plasma density, it assumes a maximal value at the coalescence point rather than at the lower hybrid resonance as in Fig.~\ref{fig:Spectrum}.

Below we briefly review all three types of the singular points in a simple plasma in order to introduce notations required for further treatment.

\subsection{Hybrid resonances}
\label{2.4.1}

For a fixed $N_{\|}$, the resonant points are determined by the condition
    \begin{equation}
    \label{2.4.1:1}
    \mathbb{A}=0.
    \end{equation}
When it is satisfied, the larger of the two solutions \eqref{2.3:14} of the dispersion equation \eqref{2.3:13} tends to infinity, $N_{\bot+}^{2}\to\infty$.

Equation \eqref{2.4.1:1} determines the so-called \emph{plasma} or \emph {hybrid resonances}. Since it is linear with respect to the square of the total plasma frequency $\omega_{p}^{2} = \omega_{pe}^{2} + \omega_{pi}^{2} $, it has a unique solution
    \begin{equation}
    \label{2.4.1:2}
        \omega_{p,\text{res}}^{2}
        %=
        %\frac{\left(\omega^{2}-\Omega_{e}^{2}\right)\left(\omega^{2}-\Omega_{i}^{2}\right)}
        %    {\omega^{2}+\Omega_{e} \Omega_{i}}
        =
        \frac{
            \left(\omega_{ce}^{2}-\omega^{2}\right)
            \left(\omega^{2}-\omega_{ci}^{2}\right)
        }{
            \omega_{ce} \omega_{ci}-\omega^{2}
        }
    .
    \end{equation}
As a negative value of $\omega_{p}^{2}$ has no meaning, the right-hand-side of Eq.~\eqref {2.4.1:2} must be positive. It occurs in the range of frequencies
    \begin{equation}
    \label{2.4.1:3}
    \omega_{ci}<\omega<\sqrt{\omega_{ce}\omega_{ci}}
    ,
    \end{equation}
which corresponds to the \emph{lower hybrid resonance} (LHR). Another range of frequencies
    \begin{equation}
    %\label{2.4.1:4}
    \omega > \omega_{ce}
    \end{equation}
corresponds to the \emph{upper hybrid resonance}. There are no plasma resonances if $\omega <\omega_{ci}$ or
    \begin{gather}
    \label{2.4.1:5}
    \sqrt{\omega_{ce}\omega_{ci}}<\omega<\omega_{ce}
    .
    \end{gather}

Below we will say that Eq.~\eqref{2.4.1:2} determines the resonant electron density $n_{\text{res}}$, which can be expressed from Eq.~\eqref{2.4.1:2} using the definition of $\omega_{p}^{2}$. Note also that so far all the formulas were exact and, in particular, we have not neglected the small ratio $m_{e}/m_{i}$. However in the range of frequencies $\omega\sim\sqrt{\omega_{ce}\omega_{ci}}$ Eq.~\eqref{2.4.1:2} can be simplified to the following expression
    \begin{equation}
    \label{2.4.1:6}
        \omega_{pe,\text{res}}^{2}
        \approx
        \frac{\omega_{ce}^{2}\omega^{2}}
            {\omega_{ce} \omega_{ci}-\omega^{2}}
    .
    \end{equation}
Taking into account the quasi-neutrality condition \eqref{2.3.7}, it can be easily reduced to the form
    \begin{equation}
    \label{2.4.1:7}
        \frac{1}{\omega^{2}_\text{LH}}
        \approx
        \frac{1}{\omega_{pi}^{2}}
        +
        \frac{1}{\omega_{ce}\omega_{ci}}
    ,
    \end{equation}
known in the literature. It approximately determines the frequency of the lower hybrid resonance and can be found in many textbooks while an exact expression
%    \begin{gather*}
%    \omega_\text{UH}
%    =
%    \frac{1}{2}\left(
%        \omega_{ce}^{2}+\omega_{ci}^{2}+\omega_{p}^{2}
%    \right)
%    -
%    \sqrt{
%        \frac{1}{4}\left(
%            \omega_{ce}^{2}+\omega_{ci}^{2}+\omega_{p}^{2}
%        \right)^{2}
%        +
%        \omega_{p}^{2}\omega_{ce}\omega_{ci}
%        -
%        \omega_{ce}^{2}\omega_{ci}^{2}
%    }
%    \end{gather*}
for $\omega_\text{LH}$ can be readily derived from Eq.~\eqref{2.4.1:2}.

%% ====================================================
%    \begin{gather*}
%    \label{2.4.1:7a}
%        \omega ^{2}
%        \approx
%        \omega _{ci}^{2} + \frac{\omega_{ci}}{\omega _{ce}}\omega _{pe}^{2}
%        =
%        \omega _{ci}^{2} + \omega _{pi}^{2}
%        .
%    \end{gather*}
%    \begin{equation}
%    \label{2.4.1:7b}
%        \frac{1}{\omega^{2}_\text{LH}}
%        \approx
%        \frac{1}{\omega_{pi}^{2}+\omega_{ci}^{2}}
%        +
%        \frac{1}{\omega_{ce}\omega_{ci}}
%    .
%    \end{equation}
%
%
%    \begin{equation}
%    \label{2.4.1:8}
%        \omega_\text{UH}
%        \approx
%        \sqrt{\omega_{pe}^{2} + \omega_{ce}^{2}}
%    .
%    \end{equation}
%
%% ====================================================

\subsection{Cutoffs}
\label{2.4.2}

The cutoff points $N_{\bot}^{2}=0$ are found from the equation
    \begin{equation}
    %\label{2.4.2:1}
    \mathbb{C} = 0
    .
    \end{equation}
Since
    \begin{equation*}
    %\label{2.4.2:2}
    \mathbb{C}
    =
    \eta
    \left(N_{\|}^{2}-\varepsilon_{+}\right)
    \left(N_{\|}^{2}-\varepsilon_{-}\right)
    ,
    \end{equation*}
the first of the cutoffs is found from the equation $\eta=0$. It occurs at the density such that
    \begin{equation}
    \label{2.4.2:3}
    \omega_{p}^{2} = \omega^{2}
    =
    \omega_{p,\text{cut}1}^{2}
    .
    \end{equation}
This density is usually very low. For example, for the frequency $\omega /2\pi = 13.56\;\text{MHz}$, which is often used in the helicon plasma sources, it is as low as $n_{e}=2.28\times10^{6}\,\text{cm}^{-3}$.
%; it is still relatively low at $\omega /2\pi = 2.45\;\text{GHz}$; then $n_{e}=1.04\times10^{10}\,\text{cm}^{-3}$.

Two more cutoffs are found by equating $N_{\|}^{2}$ to $\varepsilon_{\pm}$, which yields
    \begin{equation*}
    \label{2.4.2:4}
    \omega_{p}^{2} =
        \left(1-N_{\|}^{2}\right)
        %\left(\omega\pm\Omega_{e}\right)
        %\left(\omega\pm\Omega_{i}\right)
        \left(\omega\pm\omega_{ce}\right)
        \left(\omega\mp\omega_{ci}\right)
        .
    \end{equation*}
For any sign of the factor $(1-N_{\|}^{2})$ only one of them hits the interval $\omega_{ci}<\omega<\sqrt{\omega_{ce}\omega_{ci}}$. In case of slow wave with $N_{\|}^{2}>1$ and $\omega_{ci} < \omega < \omega_{ce}$,
    \begin{equation}
    \label{2.4.2:4a}
    \omega_{p}^{2} =
        \left(N_{\|}^{2}-1\right)
        %\left(\omega\pm\Omega_{e}\right)
        %\left(\omega\pm\Omega_{i}\right)
        \left(\omega_{ce}-\omega\right)
        \left(\omega+\omega_{ci}\right)
    =
    \omega_{p,\text{cut}2}^{2}
        .
    \end{equation}
%
%If $N_{\|}^{2} \gg 1$, this second cutoff corresponds to the plasma density such that
%    \begin{equation}
%    \label{2.4.2:5}
%    \omega_{pe}^{2} \approx  \left(\omega_{ce}-\omega\right) \omega N_{\|}^{2}
%    .
%    \end{equation}
%
In a typical helicon plasma source, the first cutoff \eqref{2.4.2:3} falls on the periphery of a plasma column, where the density is low, while the second cutoff \eqref{2.4.2:4a} is located in a more dense plasma core.
%In other words, the first cutoff point is located to the left from  the second point at a density scale.

\subsection{Coalescence points}
\label{2.4.3}

The monographs on plasma physics usually describe the wave dispersion in a cold magnetized plasma at a fixed angle of propagation $\theta = \arccos(N_{\|}/N)$. In this case, as seen in Fig.~\ref{fig:Spectrum}, the dispersion curves are monotonically rising functions of the wavenumber $k$, they do not intersect and do not touch each other (except for the case $\theta = 0$). However, at a fixed value of $N_{\|}$, there appear the points of coalescence, where the roots of Eq.~\eqref{2.3:13} merge as shown in Fig.~\ref{fig:Spectrum-2}. The coalescence points are found from the equation
    \begin{equation}
    \label{2.5:1}
    \mathbb{B}^{2} - 4\mathbb{A}\mathbb{C} =0
    ,
    \end{equation}
which means that $N_{\bot+}^{2}=N_{\bot-}^{2}$. Equation \eqref{2.5:1} is quadratic with respect to $\omega_{p}^{2} $ and, therefore, has two solutions, which we denote as $\omega_{p,\text{coal}1}^{2}$ and $\omega_{p,\text{coal}2}^{2}$; we do not give here explicit expressions for these quantities because they are too complex.
%, assuming that $\omega_{p,\text{coal}1}^{2}\leq \omega_{p,\text{coal}2}^{2}$.

\begin{figure}\centering
    \includegraphics[width=\columnwidth]{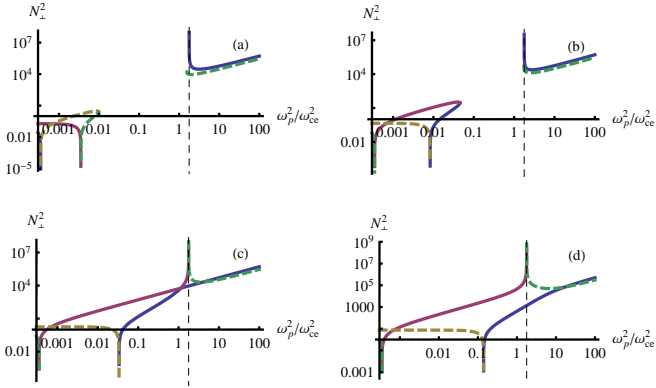}
    \caption{
        (Color online)
        To the derivation of the Golant-Stix criterion. The solid line shows $N_{\bot}^{2}$ for a propagating solution $N_{\bot}^{2}>0$ of the dispersion equation \eqref{2.3:13}, and the dashed line shows $-N_{\bot}^{2}$ for an evanescent solution $N_{\bot}^{2}<0$; $\omega/\sqrt{\omega_{ci}\omega_{ce}}=0.8$, $\omega_{ce}/\omega_{ci}=1836.15$,
        (a) $N_{\|}=0.9$,
        (b) $N_{\|}=1.2$,
        (c) $N_{\|}=5/3$,
        (d) $N_{\|}=2.9$.
        %Merge point roots of the dispersion equation lie at the intersection of the blue and purple curves. Gap on the top two graphs between a pair of points where there is no real solutions , corresponding to the second zone opacity of the plasma at moderate densities , $ n_{\text{coal}1} < n <n_{\text{coal}2}$; first opacity zone occurs at low density $n < n_\text{P} $ (where $\omega_{p} <\omega$). The second zone opacity disappears for sufficiently large longitudinal refractive index $ N_{\|}> N_{\|, \text{crit}}$.
        The vertical dashed line indicates the position of the lower hybrid resonance.
        }
    \label{fig:1}
\end{figure}
A solution of the dispersion equation \eqref{2.3:13} for several values of the refractive index  $N_{\|}$ and a fixed frequency $\omega$ is shown in Fig.~\ref{fig:1} depending on the dimensionless parameter $\omega_{p}^{2}/\omega_{ce}^{2}$.
Figures~\ref{fig:1},a and ~\ref{fig:1},b show that at a relatively low value of $N_{\|}$ a zone of opacity, where $N_{\bot}^{2}$ is negative or complex for both branches \eqref{2.3:14}, is located between the two zones of transparency, where $N_{\bot}^{2}>0$ for at least one of the two branches. The lower hybrid resonance is located in the upper density zone of transparency. The opaque zone makes it inaccessible for a wave propagating inward radially from outside of the plasma column. A low density zone of transparency matches a vacuum region (where $\omega_{p}^{2}/\omega_{ce}^{2}=0$) if $N_{\|}^{2}<1$  (Fig.~\ref{fig:1},a). However, the vacuum region becomes opaque if $N_{\|}^{2}>1$  (Fig.~\ref{fig:1},b); in this case, the low density zone of transparency begins from the first cutoff \eqref{2.4.2:3}.
As $N_{\|}$ increases, the opaque zone, located between the low density and the high density zones of transparency, gradually shrinks. Its boundaries are the points of coalescence, where $\omega_{p}^{2}$ is either equal to $\omega_{p,\text{coal}1}^{2}$ or $\omega_{p,\text{coal}2}^{2}$.

%At its edges are zones of transparency that can be distributed simultaneously both branches of oscillations (Fig.~\ref {fig:1}, b). These branches meet two positive roots \ eqref {2.3:14} equation \ eqref {2.3:13}. Area on the border with opacity they merge at the points of coalescence, in each of which $ N_ {\bot+}^{2} = N_{\bot-}^{2}$, ie \, is
%    \begin{equation}
%    \label{2.5:1}
%    b^{2} - 4ac =0
%    .
%    \end{equation}

\section{The Golant-Stix criterion}
\label{s4}

The opaque zone, described in Sec.~\ref{2.4.3}, disappears when
    \begin{gather}
    \label{2.5:1a}
        \omega_{p,\text{coal}1}^{2}=\omega_{p,\text{coal}2}^{2}
    \end{gather}
and the two coalescence points merge as shown in Fig.~\ref{fig:1},c. Equation \eqref{2.5:1a} has a unique solution with respect to $N_{\|}^{2}$ and determines a critical value
    \begin{equation}
    \label{2.5:2}
    N_{\|,\text{crit}}^{2}
    =
    \frac{\omega_{ce}\omega_{ci}}{\omega_{ce}\omega_{ci}-\omega^{2}}
    \end{equation}
of the longitudinal refractive index. The opaque zone is absent and, hence, the lower hybrid resonance is accessible, if
    \begin{equation}
    \label{2.5:2a}
    N_{\|}^{2}
    \geqslant
    N_{\|,\text{crit}}^{2}
    .
    \end{equation}
Note that the expression \eqref{2.5:2} is exact and obtained without any simplifying assumption. Although it is rather simple, its derivation is somewhat cumbersome and was performed using the Wolfram Mathematica \cite{WolframMathematica}.

The condition \eqref{2.5:2a} is equivalent to the Golant-Stix criterion, which specifies the conditions of the penetration into the plasma of an electromagnetic wave with a frequency of the order of the lower hybrid resonance frequency. The criterion was previously derived by V.E. Golant in
%
%Refs.~\cite{Golant1971ZhTPh_41_2492, Golant1972SovPhysTechPhys_16_1980}
Ref.~\cite{Golant1972SovPhysTechPhys_16_1980}.
He discarded some small terms
and wrote his criterion in the form
    \begin{equation}
    \label{2.5:3}
    N_{\|}^{2}
    \geqslant
    1 + \frac{\omega_{pe,\text{res}}^{2}}{\omega_{ce}^{2}}
    \end{equation}
(see Eq.~(12) in \cite{Golant1972SovPhysTechPhys_16_1980}), where the resonant value of the electron plasma frequency $\omega_{pe,\text{res}}^{2}$ was determined from the approximate equation
    \begin{equation*}
    %\label{2.5:4}
    1
    - \frac{\omega_{pe}^{2}}{\omega_{ce}^{2}}
    + \frac{\omega_{pi}^{2}}{\omega^{2}}
    =
    1
    +
    \frac{\omega_{pe}^{2}}{\omega_{pe,\text{res}}^{2}}
    .
    \end{equation*}
Its solution
%    \begin{equation}
%    \label{2.5:5}
%    \omega_{pe,\text{res}}^{2}
%    %=
%    %\frac{-\Omega_{e}^{2} \omega^{2}}{\omega^{2}+\Omega_{e} \Omega_{i}}
%    =
%    \frac{\omega_{ce}^{2} \omega^{2}}{\omega_{ce} \omega_{ci}-\omega^{2}}
%    \end{equation}
coincides with approximate Eq.~\eqref{2.4.1:6}. Surprisingly, but the substitution of the approximate expression \eqref{2.4.1:6} to approximate inequality \eqref{2.5:3} recovers the exact criterion \eqref{2.5:2}.

A subtle derivation of a criterion similar to \eqref{2.5:3} can be found in the textbook  \cite{Stix1992} (see \S4-12 and Eq.~(103) there). For this reason, the authorship of the criterion \eqref{2.5:3} is also attributed to R.H. Stix.

%%=============================================
%    \begin{gather}
%    \label{2.5:6}
%    N_{\|}
%    >
%    \sqrt{
%        1+\frac{\omega_{pe}^{2}}{\omega_{ce}^{2}}
%        \left(
%            1-\frac{\omega_{ce}\omega_{ci}}{\omega^{2}}
%        \right)
%    }
%    +
%    \frac{\omega_{pe}}{\omega_{ce}}
%    .
%    \end{gather}
%    \begin{gather}
%    \label{2.5:6a}
%    N_{\|}
%    >
%    \sqrt{
%        1
%        -\frac{\omega_{pi}^{2}}{\omega^{2}}
%        +
%        \frac{\omega_{pe}^{2}}{\omega_{ce}^{2}}
%    }
%    +
%    \frac{\omega_{pe}}{\omega_{ce}}
%    \end{gather}
%    \begin{gather*}
%        \omega _{pe}^{2} =
%        \frac{
%            \omega_{ce}^{2}\omega^{2}
%        }{
%            \omega_{ce}\omega_{ci}-\omega^{2}
%        }
%    \end{gather*}

Before concluding this section, it should be emphasized that the Golant-Stix criterion refers to the case when $\omega <\sqrt{\omega_{ce}\omega_{ci}}$, so that a lower hybrid resonance can exist in the plasma column. As is clear from Eq.~\eqref{2.5:2}, the critical value $N_{\|,\text{crit}}^{2}$ of the square of the refractive index formally becomes negative if $\omega > \sqrt{\omega_{ce}\omega_{ci}}$. It means that the opaque zone cannot shrink to zero and, hence, the high density transparent zone is inaccessible in if  $\omega >\sqrt{\omega_{ce}\omega_{ci}}$. The high-frequency regime of the helicon plasma sources operation is considered in Sec.~\ref{s7}.

\section{Limiting density}
\label{s5}

Equation \eqref{2.5:2} determines a critical value of the longitudinal wave number $k_{\|,\text{crit}} = \left(\omega/c\right) N_{\|,\text{crit}}$. Since $N_{\|,\text{crit}} > 1$, an antenna must launch a slowed wave. The slower the wave, the higher the density of the plasma that can be heated. Indeed,
    \begin{equation}
    \label{2.6:02}
    \omega_{p,\text{res}}^{2}
    =
    \frac{\left(\omega_{ce}^{2}-\omega^{2}\right)\left(\omega^{2}-\omega_{ci}^{2}\right)}
    {\omega_{ce} \omega_{ci}-\omega^{2}}
    \approx
    %\frac{m_{i}}{Zm_{e}}
    \frac{\omega_{ce}}{\omega_{ci}}
    k_{\parallel,\text{crit}}^{2}c^{2}
    .
    \end{equation}
In practical units, the resonant electron density is
    \begin{equation}
    \label{2.6:03}
    n_{e,\text{res}}\, [\text{cm}^{-3}]
    =2\times10^{16}AZ^{-1}(\lambda_{\|}\,[\text{cm}])^{-2}
    ,
    \end{equation}
where $A$ it the atomic weight of the plasma ions, $Z$ is their charge state, $n_{e}$ is expressed in $\text{cm}^{-3}$, and the wave length $\lambda_{\|}=2\pi/k_{\|}$ in $\text{cm}$.

In the vacuum region the wave is evanescent since
    \begin{gather*}
    %\label{2.6:06}
    N_{\bot,\text{vac}}^{2}=1-N_{\|,\text{crit}}^{2}
    =
    \frac{-\omega^{2}}{\omega_{ce}\omega_{ci}-\omega^{2}}
%    \\
%    =
%    \frac{\omega_{p,\text{res}}^{2}\omega^{2}}{
%        \left(\omega^{2}-\omega_{ce}^{2}\right)
%        \left(\omega^{2}-\omega_{ci}^{2}\right)
%    }
    \approx
    -\frac{\omega_{p,\text{res}}^{2}}{\omega_{ce}^{2}}
    <0
    .
    \end{gather*}
Therefore, for achieving a higher density an antenna must be placed as closer to the plasma as possible in order to reduce the wave attenuation in the opaque area at the plasma periphery; this might be a difficult technical problem.

Rarefied periphery of the plasma is also opaque to the wave up to the first cutoff, i.e. at
    \begin{equation*}
    %\label{2.6:04}
    \omega_{p}^{2}<\omega^{2}
    .
    \end{equation*}
The cutoff density is substantially smaller than the resonant one since
    \[
    %\label{2.6:05}
    \frac{\omega^{2}}{\omega_{p,\text{res}}^{2}}
    \approx
    %\frac{Z m_{e}}{m_{i}}\,
    \frac{\omega_{ci}}{\omega_{ce}}\,
    N_{\parallel,\text{crit}}^{-2}
    <\frac{Z m_{e}}{m_{i}}
    .
    \]
Therefore one can hope that external rf field can tunnel through the peripheral opaque zone to the plasma core.

Equations \eqref{2.5:2} and \eqref{2.6:02} in principle solve the problem of optimization of a helicon plasma source at a given frequency of the RF field and a given magnetic field. Equation \eqref{2.5:2} yields the required wavelength, and Eq.~\eqref {2.6:02} gives the maximal density of the plasma, which can be heated at such parameters.

\section{Optimal magnetic field}
\label{s6}

We can change a statement of the problem to begin with, so to speak, the antenna. Suppose that the frequency and wavelength are fixed by the antenna system design;  it means that $N_{\|} = k_{\|}c/\omega$ is a given parameter. Rewriting  Eq.~\eqref{2.5:2} in the form
    \begin{gather}
    \label{2.7:01}
        \omega _{ci}\omega_{ce}=
        \omega ^{2}
        \frac{N_{\|}^{2}}{N_{\|}^{2}-1}
    \end{gather}
then determines the magnitude of the magnetic field. In practical units,
    \begin{gather}
    \label{2.7:02}
    B_{\ast}\, [\text{kG}]
    =
    %15.3078\,
    15.3\,
    \sqrt{\frac{A}{Z}}
    \frac{N_{\|}}{\sqrt{N_{\|}^{2}-1}} f\, [\text{GHz}]
    ,
    \end{gather}
where $f=\omega /2\pi$ is the linear frequency, expressed in gigahertz. For $f=13.56\,\text{MHz}$ and $N_{\|}^{2}\gg 1$ it yields  $B_{\ast}=208\,\text{G}$.

%    \begin{gather*}
%        N_{\|}^{2}
%        =
%        1
%        -
%        \frac{\omega _{p}^{2}}{
%        \left(
%            \omega -\omega _{ce}
%        \right)
%        \left(
%            \omega +\omega _{ci}
%        \right)
%        }
%        .
%    \end{gather*}

Meaning of $B_{\ast}$ can be understood as follows. If $B<B_{\ast}$, the lower hybrid resonance is separated from the low density zone of transparency by the opaque zone where $\omega_{p,\text{coal}1}^{2} < \omega _{p}^{2} < \omega_{p,\text{coal}2}^{2}$. In this case, the helicon and TG waves merge at $\omega_{p}^{2} = \omega_{p,\text{coal}1}^{2}$.
Conversely, when $B>B_{\ast}$ TG branch propagates till the lower hybrid resonance at $\omega_{p}^{2}=\omega_{p,\text{res}}^{2}$, and the helicon branch can penetrate into even more dense plasma where it can deposit energy due to particle collisions.

Thus, $B_{\ast}$ is a minimal magnetic field required to switch on the mechanism of plasma heating due to lower hybrid resonance. Some experiments (see in particular \cite{Mori+2004PSSciTechn_13_424, Dudnikov2012moppd047, Dudnikov+2012RUPAC}) demonstrate that helicon discharge is most easily fired when $B\sim B_{\ast}$ in that sense that required rf power supply is minimal. However, the helicon plasma sources are known to effectively operate even at smaller magnetic field. In other words, a frequency in the range \eqref{2.4.1:5} can also be effectively used in such sources. We proceed to the analysis of this range in the next Section.

%    \begin{gather*}
%    %\label{2.7:03}
%        \omega _{p,\text{res}}^{2}
%        =
%        \frac{
%            \left(
%                N_{\|}^{2}\left(1-\mu\right)-1
%            \right)
%            \left(
%                N_{\|}^{2}\left(1-\mu\right)+\mu
%            \right)
%        }{
%            \left(
%                N_{\|}^{2}-1
%            \right)
%            \mu
%        }
%        \omega ^{2}
%        ,
%    \end{gather*}
%    \begin{gather*}
%    %\label{2.7:04}
%        \omega_{p,\text{res}}^{2}
%        \approx
%        \frac{m_{i}}{m_{e}Z}
%        k_{\|}^{2}c^{2}
%        ,
%    \end{gather*}

\section{High-frequency helicon sources}
\label{s7}

\begin{figure}[!tbhp]
  \centering
  \includegraphics[width=\columnwidth]{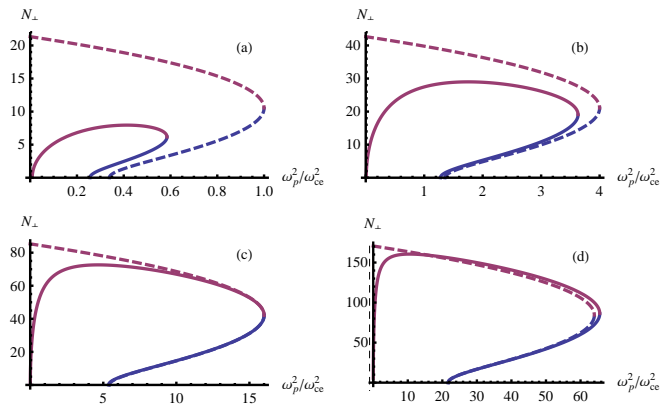}\\
  \caption{
    (Color online)
    To the derivation of the Shamrai-Taranov criterion.  Solid line shows an exact solution of Eq.\eqref{2.3:13}, and dashed line is approximate solution \eqref{6.5:3}. Accuracy of the approximation improves as $N_{\|}^{2}$ grows;
    $\omega/\sqrt{\omega_{ci}\omega_{ce}}=4$:
    (a) $N_{\|}=2$,
    (b) $N_{\|}=4$,
    (c) $N_{\|}=8$,
    (d) $N_{\|}=16$.
    Blue and purple curves are respectively the helicon and TG waves.
  }\label{fig:LHW5}
\end{figure}
For a frequency in the range \eqref{2.4.1:5}, the helicon and TG waves can propagate in the low density zone of transparency as shown in Fig.~\ref{fig:LHW5}. In this case, the maximal plasma density, that can be heated by these waves at given $\omega $ and $k_{\|}$, is defined by
    \begin{gather}
    \omega^{2}=\omega_{p,\text{coal}1}^{2}
    .
    \end{gather}
It is limited by the coalescence of the helicon and TG waves. The effect of coalescence can be understood using the Appleton-Hartree-Booker simplified dispersion relation of the helicon waves (see eg. \cite{Chen173, Akhiezer+1974(eng)})
    \begin{gather}
    \label{6.5:1}
%    \frac{1}{N_{\|}}
%    =
%    \frac{\omega_{ce} kc}{\omega_{pe}^{2}+k^{2}c^{2}}
%    \\
    \omega
    =
    \frac{\omega_{ce} kk_{\|}c^{2}}{\omega_{pe}^{2}+k^{2}c^{2}}
    ,
    \end{gather}
where $k=\sqrt{k_{\bot}^{2}+k_{\|}^{2}}$. In the limit $k_{\bot}\to\infty $ it also describes TG waves.

\begin{figure}[!tbhp]
  \centering
  % See Shamrai-Taranov.nb
  \includegraphics[width=0.8\columnwidth]{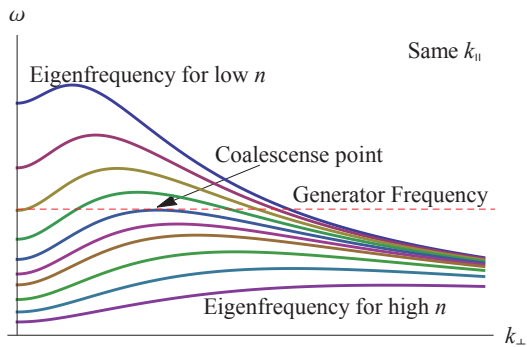}\\
  \caption{
    (Color online)
    Eigenfrequency of the helicon and TG waves vs. $k_{\bot}$ for different values of the plasma density and a fixed $k_{\|}$.
  }\label{fig:Shamrai-Taranov-1}
\end{figure}
The eigenfrequencies \eqref{6.5:1} are drawn in Fig.~\ref{fig:Shamrai-Taranov-1} for various values of $\omega_{pe}$ at a fixed value of $k_{\|}$.  The upper curves correspond to the periphery of the plasma column where $\omega_{pe}^{2}$ is small whereas the bottom curves represent the column core where $\omega_{pe}^{2}$ is larger. A horizontal line represents frequency $\omega $ of the antenna so that propagating waves correspond to the intersection point of the solid curves with this line.  The horizontal line crosses the upper curves only once (eg., blue upper curve in the figure), which means that only one branch of the waves can propagate at the plasma periphery, namely the TG wave. A second intersection point with smaller $k_{\bot}$, which corresponds to the helicon wave, appears closer to the center plasma (dark-yellow curve). Even closer to the plasma column core, the dispersion curves pass below the horizontal line, which means that a sufficiently dense plasma is not transparent to the waves with given $\omega$ and $k_{\|}$. This conclusion is qualitatively confirmed by Fig.~\ref{fig:LHW5} which draws both exact and approximate solutions of the dispersion equation.

Equation \eqref{6.5:1}, rewritten in the form
    \begin{equation}
    \label{6.5:2}
    k^{2}c^{2} - \omega_{ce}N_{\|}kc+\omega_{pe}^{2}=0
    ,
    \end{equation}
has a solution
    \begin{equation}
    \label{6.5:3}
    k_{\pm}c = \tfrac{1}{2}\omega_{ce}N_{\|}
        \pm
        \sqrt{
            \tfrac{1}{4}\omega_{ce}^{2}N_{\|}^{2}
            -
            \omega_{pe}^{2}
        }
    .
    \end{equation}
It is real, and therefore describes two propagating waves, when
    \begin{equation}
    \label{6.5:4}
    \omega_{pe}^{2}
    <
    \tfrac{1}{4}\omega_{ce}^{2}N_{\|}^{2} = \omega_{pe,\max}^{2}
    .
    \end{equation}
The condition \eqref{6.5:4} poses the density limit in the high-frequency regime of operation of a helicon plasma source. In practical units
    \begin{gather}
    n_{e,\text{max}}\, [\text{cm}^{-3}]
    =
    0.5\times10^{16}AZ^{-1}(\omega_{ci}\omega_{ce}/\omega^{2})(\lambda_{\|}\,[\text{cm}])^{-2}
    .
    \end{gather}
The condition \eqref{6.5:4} was first obtained by K. P. Shamrai and V. B. Taranov  in Ref.~\cite{ShamraiTaranov1996PSSciTech_5_474}. In comparison with the accurate criterion
    \begin{gather}
    \omega_{p}^{2} < \omega_{p,\text{coal}1}^{2}
    ,
    \end{gather}
it provides a reasonable accuracy only if $N_{\|}^{2}\gg 1$. Indeed, as seen from Fig.~\ref{fig:LHW5}, Eq.~\eqref{6.5:3} becomes accurate at sufficiently large $N_{\|}$.

A  dense plasma, $\omega_{pe}^{2} > \omega_{pe,\max}^{2}$, is opaque for both branches of the waves (the helicons and TG modes) since $k_{\bot\pm}^{2}$ are complex there. In a non-uniform plasma, merging (degeneration) of the two different wave branches is expected to be followed by the mutual linear conversion of these waves (see e.g. \cite{Stix1992}). Thus, a helicon wave converts into a TG wave near the surface $\omega_{pe} = \omega_{\max}$ and vice versa.
Note that the smaller of the two wave branches (\ref{6.5:3}), which corresponds to the minus sign, is the helicon wave; the second, signed by $+$, is the TG wave. The helicons also have a lower limit on the density defined by $k_{\bot}=0$, ie, $k=k_{\|}$ so the rarefied plasmas are opaque for helicons as seen in Fig.~\ref{fig:LHW5}.  Substituting $k=k_{\|}$ in \eqref{6.5:2} yields the cutoff plasma frequency
    \begin{equation}
    \label{6.5:5}
    \omega_{pe,\min}^{2} = \omega\left(\omega_{ce} - \omega\right)N_{\|}^{2}
    .
    \end{equation}
Thus, the helicon wave can propagate in the finite density range
    \begin{equation}
    \label{6.5:6}
    \omega_{pe,\min}^{2} <
    \omega_{pe}^{2}
    <
    \omega_{pe,\max}^{2}
    .
    \end{equation}
It shrinks to zero width at $\omega=\tfrac{1}{2}\omega_{ce}$.

%The condition for TG wave propagation is more lenient, $\alpha < 1/4$, or $n < n_{\max}$. Note that an analysis of conditions for wave propagation based on the exact fourth-order equation for the refractive indices was presented in [[13]].

\section{Discussion}
\label{s8}

We have derived two expressions for the limiting plasma density in a helicon plasma source. According to Eq.~\eqref{2.6:02}, obtained for the case $\omega <\sqrt{\omega_{ce}\omega_{ci}}$, at a given $k_{\|}$ maximal density is defined by the relation
    \begin{equation}
    \label{6.8:01}
    \omega_{pe,\max}^{2}
    \approx
    \frac{\omega_{ce}}{\omega_{ci}}
    k_{\parallel}^{2}c^{2}
    .
    \end{equation}
%assuming that $k_{\|}^{2}c^{2}\gg\omega^{2}$
In case $\omega >\sqrt{\omega_{ce}\omega_{ci}}$, according to Eq.~\eqref{6.5:4}
\begin{equation}
    \label{6.8:02}
    \omega_{pe,\max}^{2}
    \approx
    \frac{\omega_{ce}^{2}}{4\omega^{2}}k_{\|}^{2}c^{2}
    .
    \end{equation}
%provided that $k_{\|}^{2}c^{2}\gg\omega^{2}$.
The expressions \eqref{6.8:01} and \eqref{6.8:02} match each other at $\omega =\frac{1}{2}\sqrt{\omega_{ce}\omega_{ci}}$ and, thus, provide a smooth objective function for choosing optimal parameters of a helicon plasma source. If we assume that $\omega$ and $k_{\|}$ are fixed by rf system design, the only remaining parameter will be the magnetic field $B$.  Starting an optimization procedure from a low magnetic field, we see that, according to Eq.~\eqref{6.8:02}, increasing $B$ would increase allowed plasma density. However, increasing $B$ above the limit $\omega = \sqrt{\omega_{ce}\omega_{ci}}$ reverts the density scaling from Eq.~\eqref{6.8:02}  to Eq.~\eqref{6.8:01}. Then, the limiting density becomes insensitive to $B$. From this reasoning, we see that the condition $\omega \approx \sqrt{\omega_{ce}\omega_{ci}}$ might be optimal for the operation of a helicon source as we suggested in Sec.~\ref{s7}. Some experiments  \cite{Mori+2004PSSciTechn_13_424, Dudnikov2012moppd047, Dudnikov+2012RUPAC} indicate that a helicon discharge is fired at a minimal rf power provided that magnetic field is within some range around the value given by Eq.~\eqref{2.7:02} although the range of allowed magnetic fields widens as rf power increases.

To the best of our knowledge, the density limits given by Eqs.~\eqref{6.8:01} and~\eqref{6.8:02} are not achieved in existing helicon plasma sources. However the results instantly grow with the increase of applied rf power so one may suppose that these limits might be met in the future.

%\bibliographystyle{apsrev4-1}
%\bibliography{LHW}

\begin{thebibliography}{18}%
\makeatletter
\providecommand \@ifxundefined [1]{%
 \@ifx{#1\undefined}
}%
\providecommand \@ifnum [1]{%
 \ifnum #1\expandafter \@firstoftwo
 \else \expandafter \@secondoftwo
 \fi
}%
\providecommand \@ifx [1]{%
 \ifx #1\expandafter \@firstoftwo
 \else \expandafter \@secondoftwo
 \fi
}%
\providecommand \natexlab [1]{#1}%
\providecommand \enquote  [1]{``#1''}%
\providecommand \bibnamefont  [1]{#1}%
\providecommand \bibfnamefont [1]{#1}%
\providecommand \citenamefont [1]{#1}%
\providecommand \href@noop [0]{\@secondoftwo}%
\providecommand \href [0]{\begingroup \@sanitize@url \@href}%
\providecommand \@href[1]{\@@startlink{#1}\@@href}%
\providecommand \@@href[1]{\endgroup#1\@@endlink}%
\providecommand \@sanitize@url [0]{\catcode `\\12\catcode `\$12\catcode
  `\&12\catcode `\#12\catcode `\^12\catcode `\_12\catcode `\%12\relax}%
\providecommand \@@startlink[1]{}%
\providecommand \@@endlink[0]{}%
\providecommand \url  [0]{\begingroup\@sanitize@url \@url }%
\providecommand \@url [1]{\endgroup\@href {#1}{\urlprefix }}%
\providecommand \urlprefix  [0]{URL }%
\providecommand \Eprint [0]{\href }%
\providecommand \doibase [0]{http://dx.doi.org/}%
\providecommand \selectlanguage [0]{\@gobble}%
\providecommand \bibinfo  [0]{\@secondoftwo}%
\providecommand \bibfield  [0]{\@secondoftwo}%
\providecommand \translation [1]{[#1]}%
\providecommand \BibitemOpen [0]{}%
\providecommand \bibitemStop [0]{}%
\providecommand \bibitemNoStop [0]{.\EOS\space}%
\providecommand \EOS [0]{\spacefactor3000\relax}%
\providecommand \BibitemShut  [1]{\csname bibitem#1\endcsname}%
\let\auto@bib@innerbib\@empty
%</preamble>
\bibitem [{\citenamefont {Chen}(1995)}]{Chen155}%
  \BibitemOpen
  \bibfield  {author} {\bibinfo {author} {\bibfnamefont {F.~F.}\ \bibnamefont
  {Chen}},\ }{\selectlanguage {english}\enquote {\bibinfo {title} {Helicon
  plasma sources},}\ }in\ \href@noop {} {{\selectlanguage {english}\emph
  {\bibinfo {booktitle} {High Density Plasma Sources}}}},\ \bibinfo {editor}
  {edited by\ \bibinfo {editor} {\bibfnamefont {O.~A.}\ \bibnamefont {Popov}}}\
  (\bibinfo  {publisher} {Noyes Publications},\ \bibinfo {address} {Park Ridge,
  NJ},\ \bibinfo {year} {1995})\ Chap.\ \bibinfo {chapter} {1}, pp.\
  \bibinfo {pages} {1--75}\BibitemShut {NoStop}%
\bibitem [{\citenamefont {Ellingboe}\ and\ \citenamefont
  {Boswell}(1996{\natexlab{a}})}]{Ellingboe1996PoP_3_2797}%
  \BibitemOpen
  \bibfield  {author} {\bibinfo {author} {\bibfnamefont {A.~R.}\ \bibnamefont
  {Ellingboe}}\ and\ \bibinfo {author} {\bibfnamefont {R.~W.}\ \bibnamefont
  {Boswell}},\ }\href {\doibase 10.1063/1.871713} {\bibfield  {journal}
  {\bibinfo  {journal} {Physics of Plasmas}\ }\textbf {\bibinfo {volume} {3}},\
  \bibinfo {pages} {2797} (\bibinfo {year} {1996}{\natexlab{a}})}\BibitemShut
  {NoStop}%
\bibitem [{\citenamefont {Shamrai}\ and\ \citenamefont
  {Taranov}(1996)}]{ShamraiTaranov1996PSSciTech_5_474}%
  \BibitemOpen
  \bibfield  {author} {\bibinfo {author} {\bibfnamefont {K.~P.}\ \bibnamefont
  {Shamrai}}\ and\ \bibinfo {author} {\bibfnamefont {V.~B.}\ \bibnamefont
  {Taranov}},\ }\href {\doibase 10.1088/0963-0252/5/3/015} {\bibfield
  {journal} {\bibinfo  {journal} {Plasma Sources Science and Technology}\
  }\textbf {\bibinfo {volume} {5}},\ \bibinfo {pages} {474–491} (\bibinfo
  {year} {1996})}\BibitemShut {NoStop}%
\bibitem [{\citenamefont {Lafleur}\ \emph {et~al.}(2011)\citenamefont
  {Lafleur}, \citenamefont {Charles},\ and\ \citenamefont
  {Boswell}}]{Lafleur+2011JPhysD_44_055202}%
  \BibitemOpen
  \bibfield  {author} {\bibinfo {author} {\bibfnamefont {T.}~\bibnamefont
  {Lafleur}}, \bibinfo {author} {\bibfnamefont {C.}~\bibnamefont {Charles}}, \
  and\ \bibinfo {author} {\bibfnamefont {R.~W.}\ \bibnamefont {Boswell}},\
  }\href {http://stacks.iop.org/0022-3727/44/i=5/a=055202} {\bibfield
  {journal} {\bibinfo  {journal} {Journal of Physics D: Applied Physics}\
  }\textbf {\bibinfo {volume} {44}},\ \bibinfo {pages} {055202} (\bibinfo
  {year} {2011})}\BibitemShut {NoStop}%
\bibitem [{\citenamefont {Shinohara}\ \emph {et~al.}(2013)\citenamefont
  {Shinohara}, \citenamefont {Tanikawa}, \citenamefont {Hada}, \citenamefont
  {Funaki}, \citenamefont {Nishida}, \citenamefont {Matsuoka}, \citenamefont
  {Otsuka}, \citenamefont {Shamrai}, \citenamefont {Rudenko}, \citenamefont
  {Nakamura} \emph {et~al.}}]{Shinohara+2013FST_63_164}%
  \BibitemOpen
  \bibfield  {author} {\bibinfo {author} {\bibfnamefont {S.}~\bibnamefont
  {Shinohara}}, \bibinfo {author} {\bibfnamefont {T.}~\bibnamefont {Tanikawa}},
  \bibinfo {author} {\bibfnamefont {T.}~\bibnamefont {Hada}}, \bibinfo {author}
  {\bibfnamefont {I.}~\bibnamefont {Funaki}}, \bibinfo {author} {\bibfnamefont
  {H.}~\bibnamefont {Nishida}}, \bibinfo {author} {\bibfnamefont
  {T.}~\bibnamefont {Matsuoka}}, \bibinfo {author} {\bibfnamefont
  {F.}~\bibnamefont {Otsuka}}, \bibinfo {author} {\bibfnamefont
  {K.}~\bibnamefont {Shamrai}}, \bibinfo {author} {\bibfnamefont
  {T.}~\bibnamefont {Rudenko}}, \bibinfo {author} {\bibfnamefont
  {T.}~\bibnamefont {Nakamura}},  \emph {et~al.},\ }\href
  {http://www.ans.org/pubs/journals/fst/a_16896} {\bibfield  {journal}
  {\bibinfo  {journal} {Fusion Science and Technology}\ }\textbf {\bibinfo
  {volume} {63}},\ \bibinfo {pages} {164} (\bibinfo {year} {2013})}\BibitemShut
  {NoStop}%
\bibitem [{\citenamefont {Ellingboe}\ and\ \citenamefont
  {Boswell}(1996{\natexlab{b}})}]{EllingboeBoswell1996PoP_3_2797}%
  \BibitemOpen
  \bibfield  {author} {\bibinfo {author} {\bibfnamefont {A.~R.}\ \bibnamefont
  {Ellingboe}}\ and\ \bibinfo {author} {\bibfnamefont {R.~W.}\ \bibnamefont
  {Boswell}},\ }\href@noop {} {\bibfield  {journal} {\bibinfo  {journal}
  {Physics of Plasmas}\ }\textbf {\bibinfo {volume} {3}} (\bibinfo {year}
  {1996}{\natexlab{b}})}\BibitemShut {NoStop}%
\bibitem [{\citenamefont {Chabert}\ \emph {et~al.}(2011)\citenamefont
  {Chabert}, \citenamefont {Braithwaite},\ and\ \citenamefont
  {Braithwaite}}]{Chabert+2011physics}%
  \BibitemOpen
  \bibfield  {author} {\bibinfo {author} {\bibfnamefont {P.}~\bibnamefont
  {Chabert}}, \bibinfo {author} {\bibfnamefont {N.}~\bibnamefont
  {Braithwaite}}, \ and\ \bibinfo {author} {\bibfnamefont {N.~S.~J.}\
  \bibnamefont {Braithwaite}},\ }\href@noop {} {\emph {\bibinfo {title}
  {Physics of Radio-Frequency Plasmas}}}\ (\bibinfo  {publisher} {Cambridge
  University Press},\ \bibinfo {year} {2011})\BibitemShut {NoStop}%
\bibitem [{\citenamefont {Golant}(1972)}]{Golant1972SovPhysTechPhys_16_1980}%
  \BibitemOpen
  \bibfield  {author} {\bibinfo {author} {\bibfnamefont {V.}~\bibnamefont
  {Golant}},\ }\href@noop {} {\bibfield  {journal} {\bibinfo  {journal} {Sov.
  Phys.-Tech. Phys.}\ }\textbf {\bibinfo {volume} {16}},\ \bibinfo {pages}
  {1980} (\bibinfo {year} {1972})}\BibitemShut {NoStop}%
\bibitem [{\citenamefont {Stix}(1962)}]{Stix1962}%
  \BibitemOpen
  \bibfield  {author} {\bibinfo {author} {\bibfnamefont {T.~H.}\ \bibnamefont
  {Stix}},\ }\href@noop {} {{\selectlanguage {english}\emph {\bibinfo {title}
  {The Theory of Plasma Waves}}}}\ (\bibinfo  {publisher} {McGraw-Hill},\
  \bibinfo {address} {New York},\ \bibinfo {year} {1962})\BibitemShut {NoStop}%
\bibitem [{\citenamefont {Stix}(1992)}]{Stix1992}%
  \BibitemOpen
  \bibfield  {author} {\bibinfo {author} {\bibfnamefont {T.~H.}\ \bibnamefont
  {Stix}},\ }\href@noop {} {{\selectlanguage {english}\emph {\bibinfo {title}
  {Waves in Plasmas}}}}\ (\bibinfo  {publisher} {AIP},\ \bibinfo {address} {New
  York},\ \bibinfo {year} {1992})\BibitemShut {NoStop}%
\bibitem [{\citenamefont {Trivelpiece}\ and\ \citenamefont
  {Gould}(1959)}]{TrivelpieceGould1959JAP_30_1784}%
  \BibitemOpen
  \bibfield  {author} {\bibinfo {author} {\bibfnamefont {A.~W.}\ \bibnamefont
  {Trivelpiece}}\ and\ \bibinfo {author} {\bibfnamefont {R.~W.}\ \bibnamefont
  {Gould}},\ }\href@noop {} {\bibfield  {journal} {\bibinfo  {journal} {Journal
  of Applied Physics}\ }\textbf {\bibinfo {volume} {30}},\ \bibinfo {pages}
  {1784} (\bibinfo {year} {1959})}\BibitemShut {NoStop}%
\bibitem [{\citenamefont {Shafranov}(1967)}]{Shafranov1963VTP3(eng)}%
  \BibitemOpen
  \bibfield  {author} {\bibinfo {author} {\bibfnamefont {V.~D.}\ \bibnamefont
  {Shafranov}},\ }{\selectlanguage {english}\enquote {\bibinfo {title}
  {Electromagnetic waves in a plasma},}\ }in\ \href@noop {} {{\selectlanguage
  {english}\emph {\bibinfo {booktitle} {Reviews of plasma physics}}}},\
  Vol.~\bibinfo {volume} {3},\ \bibinfo {editor} {edited by\ \bibinfo {editor}
  {\bibfnamefont {M.~A.}\ \bibnamefont {Leontovich}}}\ (\bibinfo  {publisher}
  {Consultants Bureau},\ \bibinfo {address} {New York},\ \bibinfo {year}
  {1967})\BibitemShut {NoStop}%
\bibitem [{\citenamefont {Akhiezer}\ \emph {et~al.}(1975)\citenamefont
  {Akhiezer}, \citenamefont {Akhiezer}, \citenamefont {Polovin}, \citenamefont
  {Sitenko},\ and\ \citenamefont {Stepanov}}]{Akhiezer+1974(eng)}%
  \BibitemOpen
  \bibfield  {author} {\bibinfo {author} {\bibfnamefont {A.~I.}\ \bibnamefont
  {Akhiezer}}, \bibinfo {author} {\bibfnamefont {I.~A.}\ \bibnamefont
  {Akhiezer}}, \bibinfo {author} {\bibfnamefont {R.~V.}\ \bibnamefont
  {Polovin}}, \bibinfo {author} {\bibfnamefont {A.~G.}\ \bibnamefont
  {Sitenko}}, \ and\ \bibinfo {author} {\bibfnamefont {K.~N.}\ \bibnamefont
  {Stepanov}},\ }\href@noop {} {{\selectlanguage {english}\emph {\bibinfo
  {title} {Plasma electrodynamics}}}},\ Vol.~\bibinfo {volume} {1}\ (\bibinfo
  {publisher} {Pergamon Press Oxford},\ \bibinfo {year} {1975})\BibitemShut
  {NoStop}%
\bibitem [{Wol(2014)}]{WolframMathematica}%
  \BibitemOpen
  \href {http://www.wolfram.com/mathematica/} {{\selectlanguage
  {english}\enquote {\bibinfo {title} {Wolfram mathematica},}\ }} (\bibinfo
  {year} {2014})\BibitemShut {NoStop}%
\bibitem [{\citenamefont {Mori}\ \emph {et~al.}(2004)\citenamefont {Mori},
  \citenamefont {Nakashima}, \citenamefont {Baity}, \citenamefont {Goulding},
  \citenamefont {Carter},\ and\ \citenamefont
  {Sparks}}]{Mori+2004PSSciTechn_13_424}%
  \BibitemOpen
  \bibfield  {author} {\bibinfo {author} {\bibfnamefont {Y.}~\bibnamefont
  {Mori}}, \bibinfo {author} {\bibfnamefont {H.}~\bibnamefont {Nakashima}},
  \bibinfo {author} {\bibfnamefont {F.~W.}\ \bibnamefont {Baity}}, \bibinfo
  {author} {\bibfnamefont {R.~H.}\ \bibnamefont {Goulding}}, \bibinfo {author}
  {\bibfnamefont {M.~D.}\ \bibnamefont {Carter}}, \ and\ \bibinfo {author}
  {\bibfnamefont {D.~O.}\ \bibnamefont {Sparks}},\ }\href
  {http://stacks.iop.org/0963-0252/13/i=3/a=009} {\bibfield  {journal}
  {\bibinfo  {journal} {Plasma Sources Science and Technology}\ }\textbf
  {\bibinfo {volume} {13}},\ \bibinfo {pages} {424} (\bibinfo {year}
  {2004})}\BibitemShut {NoStop}%
\bibitem [{\citenamefont {Dudnikov}\ \emph
  {et~al.}(2012{\natexlab{a}})\citenamefont {Dudnikov}, \citenamefont
  {Johnson}, \citenamefont {Murray}, \citenamefont {Pennisi}, \citenamefont
  {Piller}, \citenamefont {Santana}, \citenamefont {Stockli},\ and\
  \citenamefont {Welton}}]{Dudnikov2012moppd047}%
  \BibitemOpen
  \bibfield  {author} {\bibinfo {author} {\bibfnamefont {V.}~\bibnamefont
  {Dudnikov}}, \bibinfo {author} {\bibfnamefont {R.~P.}\ \bibnamefont
  {Johnson}}, \bibinfo {author} {\bibfnamefont {S.}~\bibnamefont {Murray}},
  \bibinfo {author} {\bibfnamefont {T.}~\bibnamefont {Pennisi}}, \bibinfo
  {author} {\bibfnamefont {C.}~\bibnamefont {Piller}}, \bibinfo {author}
  {\bibfnamefont {M.}~\bibnamefont {Santana}}, \bibinfo {author} {\bibfnamefont
  {M.}~\bibnamefont {Stockli}}, \ and\ \bibinfo {author} {\bibfnamefont
  {R.}~\bibnamefont {Welton}},\ }in\ \href@noop {} {\emph {\bibinfo {booktitle}
  {Proceedings of IPAC2012, New Orleans, Louisiana, USA}}},\ \bibinfo {series
  and number} {\bibinfo {number} {MOPPD047}}\ (\bibinfo {address} {New Orleans,
  Louisiana, USA, 20-25 May, 2012},\ \bibinfo {year} {2012})\ pp.\ \bibinfo
  {pages} {469--471}\BibitemShut {NoStop}%
\bibitem [{\citenamefont {Dudnikov}\ \emph
  {et~al.}(2012{\natexlab{b}})\citenamefont {Dudnikov}, \citenamefont
  {Johnson}, \citenamefont {Murray}, \citenamefont {Pennisi}, \citenamefont
  {Piller}, \citenamefont {Santana}, \citenamefont {Stockli},\ and\
  \citenamefont {Welton}}]{Dudnikov+2012RUPAC}%
  \BibitemOpen
  \bibfield  {author} {\bibinfo {author} {\bibfnamefont {V.}~\bibnamefont
  {Dudnikov}}, \bibinfo {author} {\bibfnamefont {R.~P.}\ \bibnamefont
  {Johnson}}, \bibinfo {author} {\bibfnamefont {S.}~\bibnamefont {Murray}},
  \bibinfo {author} {\bibfnamefont {T.}~\bibnamefont {Pennisi}}, \bibinfo
  {author} {\bibfnamefont {C.}~\bibnamefont {Piller}}, \bibinfo {author}
  {\bibfnamefont {M.}~\bibnamefont {Santana}}, \bibinfo {author} {\bibfnamefont
  {M.}~\bibnamefont {Stockli}}, \ and\ \bibinfo {author} {\bibfnamefont
  {R.}~\bibnamefont {Welton}},\ }in\ \href
  {http://accelconf.web.cern.ch/accelconf/rupac2012/papers/tuppb058.pdf}
  {{\selectlanguage {english}\emph {\bibinfo {booktitle} {XXIII Russian
  Particle Accelerator Conference RuPAC'2012}}}}\ (\bibinfo {organization}
  {Saint-Petersburg, Russia, September, 24 - 28, 2012},\ \bibinfo {year}
  {2012})\ pp.\ \bibinfo {pages} {439--441}\BibitemShut {NoStop}%
\bibitem [{\citenamefont {Boswell}\ and\ \citenamefont {Chen}(1997)}]{Chen173}%
  \BibitemOpen
  \bibfield  {author} {\bibinfo {author} {\bibfnamefont {R.~W.}\ \bibnamefont
  {Boswell}}\ and\ \bibinfo {author} {\bibfnamefont {F.~F.}\ \bibnamefont
  {Chen}},\ }\href@noop {} {\bibfield  {journal} {\bibinfo  {journal} {{IEEE}
  Transactions On Plasma Science}\ }\textbf {\bibinfo {volume} {25}},\ \bibinfo
  {pages} {1229} (\bibinfo {year} {1997})}\BibitemShut {NoStop}%
\end{thebibliography}


\end{document}